\documentclass[conference]{IEEEtran}
\IEEEoverridecommandlockouts

\usepackage{cite}
\usepackage{amsmath,amssymb,amsfonts}
\usepackage{graphicx}
\usepackage{booktabs}
\usepackage{array}
\usepackage{textcomp}
\usepackage{xcolor}

\usepackage{subcaption}
\usepackage{xurl}

\newcommand{\BibTeX}{\textrm{B\kern-.05em\textsc{i\kern-.025em b}\kern-.08em
    T\kern-.1667em\lower.7ex\hbox{E}\kern-.125emX}}

\usepackage[absolute,overlay]{textpos}

\begin{document}

\title{When Coupling Backfires: BBR-Induced L4S Degradation under DualPI2}

\author{
\IEEEauthorblockN{Fatih Berkay Sarpkaya, Fraida Fund, and Shivendra Panwar}
\IEEEauthorblockA{
\textit{NYU Tandon School of Engineering}\\
Brooklyn, NY, USA\\
fbs6417@nyu.edu, ffund@nyu.edu, panwar@nyu.edu
}
}

\maketitle

\begin{abstract}
Low Latency, Low Loss, Scalable Throughput (L4S) uses DualPI2 to separate low latency (e.g., TCP Prague) and Classic flows into two queues while coupling their congestion signals. However, this coupling is designed for Reno-like Classic congestion controls, while widely deployed BBR behaves differently. Prior studies observed severe Prague throughput degradation when BBR occupies the Classic queue under DualPI2, but the mechanism remains insufficiently explained. Through controlled testbed experiments and an extension of an existing BBRv1/BBRv2 fluid model with PI2, we show that BBR can drive the PI2 base probability $\pmb{\boldsymbol{p'}}$ far above CUBIC levels, producing strong coupled marking that can nearly starve Prague despite low L4S queueing delay. The model explains BBRv1's persistent high-$\pmb{\boldsymbol{p'}}$ regime, BBRv2's lower, oscillating $\pmb{\boldsymbol{p'}}$, and their dependence on the PI2 target relative to RTT. Additional experiments show that Prague throughput degradation also occurs with BBRv3. Our findings highlight the need for coupling mechanisms supporting low latency flows across heterogeneous Classic congestion controls.
\end{abstract}


\begin{textblock*}{0.83\paperwidth}(18mm,258mm)
\scriptsize
\noindent \textcopyright{} 2026 IEEE. Personal use of this material is permitted. Permission from IEEE must be obtained for all other uses, in any current or future media, including reprinting/republishing this material for advertising or promotional purposes, creating new collective works, for resale or redistribution to servers or lists, or reuse of any copyrighted component of this work in other works.
\end{textblock*}

\section{Introduction}
\label{sec:introduction}

Many effective techniques for reducing Internet latency have
been proposed, but moving them from proposal to practice remains
challenging. They often require changes to end hosts, network
devices, or both, and must work alongside legacy protocols that
will remain in use for years. Low Latency, Low Loss, Scalable
Throughput (L4S) is a recent architecture designed around
these challenges~\cite{l4sarch-rfc9330,l4s-ecn-rfc9331}. It
combines scalable congestion control at end hosts with a coupled
Active Queue Management (AQM), DualPI2~\cite{dualpi-rfc9332}, that isolates L4S traffic (e.g., TCP Prague flows) into its
own queue but couples its congestion signal to that of the other queue, which carries Classic traffic. Controlled evaluations have confirmed that L4S delivers on
its promise in settings that match its design
assumptions (e.g., with CUBIC~\cite{cubic} as the Classic
traffic)~\cite{should-we-care-L4S}, and early field measurements
have shown tail-latency reductions for selected applications in
residential broadband~\cite{feamster}.

\begin{figure}[t]
    \centering
    \includegraphics[width=\columnwidth]{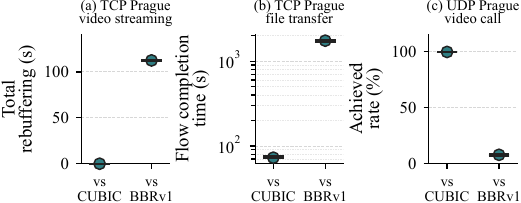}
    \caption{Application-level impact of Prague coexistence with
    Classic traffic under DualPI2. Five Prague application flows
    share a 100~Mbps bottleneck (30~ms base RTT, 15~ms target, 8-BDP buffer) with
    five CUBIC or BBRv1 bulk flows. Markers show medians over five
    trials; error bars show min--max.}
    \label{fig:application-motivation}
\end{figure}

However, not all Classic traffic matches these assumptions:
BBR, which is widely deployed~\cite{keeping-eye-on-congestion-control},
does not respond to congestion like CUBIC.
Prior studies observed severe Prague throughput degradation
with BBR in the Classic queue~\cite{L4S-PAM,anrw-L4S}, but did not fully
explain the mechanism. The effect is not limited to bulk
transfers. We ran testbed experiments
(Figure~\ref{fig:application-motivation}) in which emulated Prague applications (video
streaming and file transfer over TCP Prague, and a video call over
UDP Prague) shared a DualPI2 bottleneck with bulk CUBIC or BBRv1
flows.
Coexistence with BBRv1 substantially increases video rebuffering
and file completion time, and reduces the achieved rate of a video
call (\% of target) for Prague. This is concerning, as it occurs under
DualPI2, the coupled AQM intended to make L4S safe to deploy
alongside Classic traffic~\cite{dualpi-rfc9332}.

In this paper, we ask: \emph{how does BBR alter the congestion signal that PI2 couples into the L4S queue, and under what conditions does this harm L4S flows?} To answer it, we combine testbed measurements with analytical modeling. On the FABRIC testbed~\cite{FABRIC}, we run TCP Prague against CUBIC, BBRv1, or BBRv2 through a Linux DualPI2 bottleneck, varying capacity, base RTT, and PI2 delay target, and record Prague throughput and the router's PI2 base probability $p'$~\cite{pi2}. We then extend an existing BBRv1/BBRv2 fluid model~\cite{bbr2-fluid-model-imc-22} with PI2 to explain the $p'$ operating regimes, and connect $p'$ to Prague throughput through DualPI2 coupling and an existing Prague throughput model~\cite{prague-formulation}. We find that $p'$ is the key parameter: depending on the configuration, BBR can drive PI2 to much higher $p'$ than CUBIC, and through coupling, this elevated $p'$ causes frequent ECN marking, severely degrading Prague throughput despite low L4S queueing delay. Finally, we show experimentally that the problem persists with BBRv3,
which degrades Prague more than BBRv2 but generally less than BBRv1
(Section~\ref{sec:bbrv3}); as no comparable fluid model exists for BBRv3,
we leave its modeling to future work.

The main contributions of this work are:\footnotetext[1]{Code and data: \url{https://github.com/fatihsarpkaya/bbr-dualpi2}}
\begin{itemize}
    \item We experimentally characterize Prague coexistence with CUBIC, BBRv1, and BBRv2 under DualPI2 across bottleneck capacities, RTTs, and PI2 delay targets, showing that elevated Classic-side $p'$ tracks Prague throughput degradation even while L4S queueing delay stays low. We also show experimentally that the effect persists with the most recent BBR version, BBRv3.
    \item We extend an existing BBRv1/BBRv2 fluid model with PI2 to explain the \(p'\) operating regimes, capturing their qualitative behavior and transitions, and connect the modeled \(p'\) to Prague throughput through DualPI2 coupling and a Prague throughput model.
    \item We release our experiment and modeling scripts and data so that others can replicate and build on this work.\footnotemark[1]
\end{itemize}

\section{Background and Related Work}
\label{sec:background}

\subsection{DualPI2 Coupling}
\label{sec:background-dualpi2}

A DualPI2 bottleneck classifies packets into Classic and L4S queues using the
ECN field in the IP header~\cite{dualpi-rfc9332}. Packets carrying ECT(1) or
CE are directed to the L4S queue, while Not-ECT and ECT(0) packets go
to the Classic queue. 

The two queues apply different AQM mechanisms. The Classic queue uses PI2,
while the L4S queue uses a native low latency AQM. In the Linux DualPI2
implementation, the latter
is a step-threshold marker with a default threshold of $1$~ms. The queues share a scheduler that prioritizes L4S while reserving a minimum scheduling share (10\% by default) for Classic traffic to prevent starvation.

Rather than allocating capacity per flow, DualPI2 supports coexistence by coupling the two queues' congestion signals. It is designed for scenarios where flow queuing is infeasible for privacy, neutrality, or simplicity requirements~\cite{l4sarch-rfc9330,dualpi-rfc9332}. For a Reno-like Classic congestion control, the sending rate is approximately
inversely proportional to the square root of the congestion probability,
whereas for a scalable congestion control such as TCP Prague it is
approximately inversely proportional to the congestion probability~\cite{pi2,dualpi-rfc9332}. Both signals are derived from the PI2 controller's base probability $p'$. The congestion probability for Classic traffic, $p_C$, is
\begin{equation}
    p_C = (p')^2 ,
    \label{eq:classic-prob}
\end{equation}
while the signal coupled into the L4S queue, $p_{\mathrm{CL}}$, is
\begin{equation}
    p_{\mathrm{CL}} = \min(kp',1).
    \label{eq:coupling}
\end{equation} 
Here, $k$ is the coupling factor. The recommended value $k=2$, used in our experiments, is derived by relating steady-state rate models of Reno/CUBIC and TCP Prague while accounting for their different queueing contributions to RTT~\cite{dualpi-rfc9332}.

PI2 transforms the same base probability differently for Classic and L4S traffic. Coupling is one-way: $p'$ contributes to L4S marking, while the native L4S AQM generates no corresponding Classic signal. Linux can mark L4S packets through either PI2 coupling or the native step-threshold AQM~\cite{dualpi-rfc9332}.

Queue separation isolates L4S traffic from Classic queueing delay, but Classic congestion control still influences L4S marking through the PI2 base probability $p'$.

\subsection{Congestion Control Response}
\label{sec:background-cc}

BBR, which is widely deployed~\cite{keeping-eye-on-congestion-control}, follows a different control principle. Instead of a loss-driven congestion window (CWND) response, BBR paces and limits inflight data based on estimates of bottleneck bandwidth and propagation RTT~\cite{cardwell2016bbr}. BBRv1 can coexist poorly with conventional loss-based congestion controls, with its behavior depending on
factors such as RTT, buffer size, and competing
traffic~\cite{scholz-bbrv1-ifip-2018}. BBRv2 adds
mechanisms that respond to congestion and loss~\cite{bbr2-cubic-fairness-2020}, but its sending rate does not follow the inverse-square-root dependence on congestion probability underlying DualPI2 coupling. More recently, BBRv3 incorporates more persistent bandwidth probing~\cite{bbrv3-ietf117}. Their different congestion responses can lead to different PI2 base probabilities $p'$.

TCP Prague lies at the other end of this coupling. Prague is a scalable
congestion control that responds to frequent ECN feedback with
small CWND adjustments~\cite{prague-paper,prague-draft-rfc9330}.
Its steady-state throughput thus depends on the fraction of packets
marked with CE. An existing analytical model relates Prague throughput to
this ECN marking ratio~\cite{prague-formulation}. In
Section~\ref{sec:prague-model}, we combine this relationship with
DualPI2 coupling to map the PI2 operating point $p'$ into an estimate
of Prague throughput.

Analytical models have also been developed for BBR itself. Ware
\emph{et al.} derived a steady-state model for the coexistence of BBRv1 and
loss-based traffic~\cite{bbr-2019-ware}. Scherrer \emph{et al.}
developed a fluid model for BBRv1 and BBRv2 that captures sender dynamics,
queue evolution, and congestion feedback under FIFO and RED
bottlenecks~\cite{bbr2-fluid-model-imc-22}. We build on the latter framework
in Section~\ref{sec:model}, retaining its BBRv1 and BBRv2 sender models while
introducing a PI2 bottleneck so the model can predict the $p'$ operating
point relevant to DualPI2 coupling.

\subsection{Prior Studies of L4S--BBR Coexistence}
\label{sec:background-related}

Prior measurement studies have examined Prague coexistence with BBR under
DualPI2 and reported substantial Prague throughput degradation when BBR
occupies the Classic queue~\cite{L4S-PAM,anrw-L4S}. These results demonstrate
that queue separation alone does not guarantee good throughput coexistence
between Classic BBR and L4S traffic. However, these studies consider a
limited range of network configurations and do not systematically
characterize how the degradation varies with bottleneck capacity, RTT, PI2
delay target, and BBR version.

The mechanism connecting BBR's behavior to Prague throughput
also remains insufficiently explained. Analytical models of BBR
primarily consider FIFO/drop-tail bottlenecks or, when they model an AQM,
RED~\cite{bbr2-fluid-model-imc-22,SCHERRER-control-theoray-bbr-cubic}. Prior L4S--BBR measurements, meanwhile,
do not trace the observed Prague degradation through the complete chain from
BBR's interaction with PI2, to the resulting $p'$, to the congestion signal
coupled into the L4S queue, and finally to Prague's throughput response. To our knowledge, this work is the first to measure this chain end to end and to model it analytically.

\section{Experiment Methodology}
\label{sec:setup}

We measure Prague throughput and $p'$ while varying the Classic congestion
control (CUBIC, BBRv1, BBRv2, or BBRv3 in Section~\ref{sec:bbrv3}), bottleneck capacity, base RTT, and PI2 delay target.

\textbf{Experiment Platform:}
We use FABRIC~\cite{FABRIC}; each node is an Ubuntu~22.04 VM with 4 CPU
cores and 32~GB of RAM.

\textbf{Network Settings:}
We use a dumbbell topology in which L4S and Classic senders share a single
bottleneck router. Each experiment runs five long-lived TCP Prague flows
against five Classic flows running Linux CUBIC, BBRv1, BBRv2, or BBRv3; for comparison with the fluid model in Section~\ref{sec:model}, we also
run five Classic flows without Prague. We use base RTTs of 10, 30, and 50~ms,
bottleneck capacities of 100 and 500~Mbps, an 8-BDP buffer, and PI2 delay
targets of 5, 15, 25, 50, and 75~ms, keeping the L4S step threshold at its
1~ms default. We emulate propagation delay with \texttt{netem}, shape and
buffer the bottleneck with \texttt{tc-htb}, disable NIC offloading, and use a
1500-byte MTU on all interfaces.

\textbf{DualPI2 Configuration:}
The bottleneck router and L4S hosts run the L4STeam Linux~6.6 kernel from the
\texttt{l4steam-6.6.y} branch, which provides TCP Prague and DualPI2. Prague packets are transmitted with ECT(1). TCP ECN is disabled for all Classic senders. We
use the default DualPI2 coupling factor $k=2$ and vary only the PI2 delay
target, while keeping the remaining L4S-side configuration unchanged. CUBIC, BBRv1, and BBRv3 senders use Linux~6.4.0 from the official BBR repository, which implements BBRv1 and BBRv3 in its \texttt{v3} branch. BBRv2 senders use the \texttt{v2alpha} branch, which implements v2 in Linux~5.13.12.

\textbf{Flow Generation and Measurements:}
We generate long-lived TCP flows with \texttt{iperf3} and record aggregate
Prague and Classic throughput, plus DualPI2 statistics from the bottleneck
router: PI2 base probability $p'$, queueing delay, and CE-marking counters.
Each 180~s run is repeated three times; we report the mean over the last
60~s, averaged across trials. 

\section{Experimental Characterization}
\label{sec:characterization}

Using the experimental sweep described in Section~\ref{sec:setup}, we first
examine how the Classic congestion control affects Prague throughput, then
relate the degradation to $p'$.

\subsection{Prague Throughput Degradation}
\label{sec:characterization-throughput}

\begin{figure}[t]
    \centering
    \includegraphics[width=\columnwidth]{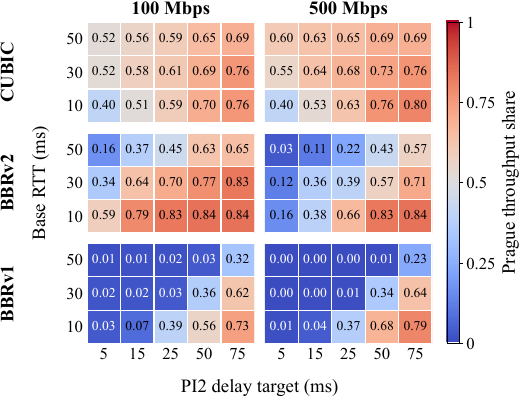}
    \caption{Normalized Prague throughput with five Prague and five CUBIC, BBRv2, or BBRv1 flows under DualPI2 (8-BDP buffer). Mean of three trials; 0.5 is an equal share. Across trials, Prague's normalized throughput
shows generally low variability, with a median standard
deviation of 0.008 across all 120 configurations, including
BBRv3 (Section~\ref{sec:bbrv3}).}
    \label{fig:prague-throughput}
\end{figure}

Figure~\ref{fig:prague-throughput} shows normalized Prague throughput. With CUBIC, Prague maintains a substantial share of
the capacity, which increases with the PI2 target (the delay target of DualPI2's PI2 controller).
Larger targets increase Classic queueing delay, giving
Prague an RTT advantage because DualPI2's fixed coupling
factor is based on representative RTTs and targets.
At low RTTs and small targets, however, Prague's
RTT-independence mechanism (using a 25~ms reference RTT)
puts it at a disadvantage relative to CUBIC.

In contrast, BBRv1 nearly starves Prague across many configurations, particularly when the PI2 target is at or below the base RTT ($T_{\mathrm{PI2}}/\mathrm{RTT} \lesssim 1$). This corresponds to a queue target of at most one BDP, consistent with BBRv1's approximately 2-BDP CWND limit. As explained in Section~\ref{sec:model-comparison}, BBRv1's insensitivity to loss is the main cause of this starvation. At a 30~ms RTT and 15~ms PI2 target, the five Prague flows obtain only about 2~Mbps in aggregate at both 100 and 500~Mbps, compared with 58 and 320~Mbps with CUBIC, respectively.

BBRv2 suppresses Prague less than BBRv1 because it reduces its inflight limit in response to loss, but it can still suppress Prague severely when the PI2 target is small relative to the base RTT, particularly at 500~Mbps. At a 30~ms RTT and 5~ms PI2 target, increasing the bottleneck capacity from 100 to 500~Mbps reduces Prague's share from 0.34 to 0.12 with BBRv2, compared with 0.52 to 0.55 with CUBIC. This stronger capacity dependence with BBRv2 is explained in Section~\ref{sec:prague-model}. With both BBR versions, Prague throughput generally recovers as the PI2 target increases. At large targets, BBR's inflight limits generally keep the queue below the target, reducing coupled marking and letting Prague obtain a larger share.

Despite these throughput differences, the L4S queue remains shallow, with p95 queueing delay at or below approximately 1.1~ms across all evaluated configurations. Thus, the coexistence problem is one of throughput, not latency.

\subsection{Classic Queue Congestion Signal}
\label{sec:characterization-p}

Having characterized Prague throughput degradation, we next examine the underlying congestion signal.

\begin{figure*}[t]
    \centering
    \includegraphics[width=0.95\textwidth]{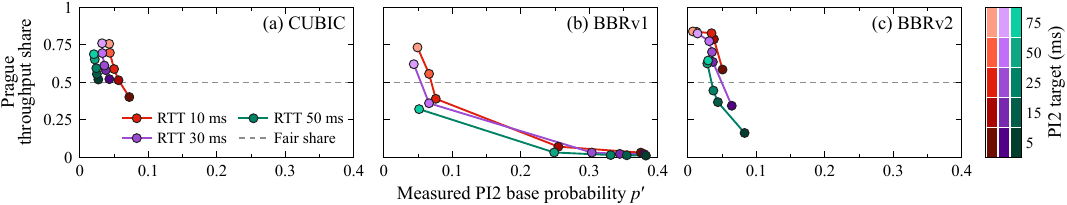}
    \caption{Higher PI2 base probability $p'$ is generally associated with lower normalized Prague throughput, with BBRv1 reaching the highest-$p'$ and most severe degradation regimes. Each trajectory varies the PI2 target (darker shades: smaller targets) at a fixed base RTT and 100~Mbps capacity; each point is the mean of three trials.}
    \label{fig:p-vs-throughput}
\end{figure*}

Figure~\ref{fig:p-vs-throughput} relates normalized Prague throughput to the router's measured mean $p'$ at 100~Mbps. At tight PI2 targets, BBRv1 drives the Classic queue to a high-$p'$ operating point, with $p'$ reaching roughly $0.25$--$0.38$ in several configurations. With $k=2$, this corresponds to a coupled marking probability of $50$--$76$\% for Prague, contributing to its near-starvation. As the PI2 target increases, $p'$ decreases and Prague recovers.

BBRv2 generally maintains lower $p'$ than BBRv1 due to its loss-responsive behavior, resulting in less severe Prague degradation. However, tighter PI2 targets still produce elevated $p'$ and reduced Prague throughput, particularly at larger RTTs. In contrast, CUBIC maintains low $p'$ across the tested configurations, avoiding the severe degradation observed with BBR.

To verify that this signal reaches Prague via coupling, we examine DualPI2's CE-marking counters. With BBRv1 at 100~Mbps, 30~ms RTT, and a 15~ms PI2 target, about 66\% of L4S packets are CE-marked, consistent with coupled marking $p_{\mathrm{CL}}=2p'\approx0.69$ ($k=2$). In this setting, all observed L4S CE marks come from PI2
coupling, none from the native L4S step marker. These results support coupled marking as the mechanism linking elevated $p'$ to Prague throughput degradation.

\section{Explaining the Classic Queue PI2 Signal}
\label{sec:model}

We now examine why different Classic congestion controls produce distinct \(p'\) operating regimes.

\subsection{Analytical Setup and PI2 Bottleneck Model}
\label{sec:model-formulation}

We extend the BBRv1/BBRv2 fluid model of~\cite{bbr2-fluid-model-imc-22} with a PI2 controller to model the base probability \(p'(t)\), while keeping its original congestion control and queue dynamics.

We consider a single bottleneck link of capacity $C$ (in segments per unit
time) and $N$ long-lived flows. Each flow $i$ has a fixed base round-trip
time $D_i$ excluding queueing, and its instantaneous round-trip time is
\begin{equation}
    \tau_i(t)=D_i+\frac{q(t)}{C},
\end{equation}
where $q(t)$ is the bottleneck queue backlog. Let $x_i(t)$ denote the sending
rate of flow $i$ and $y(t)=\sum_i x_i(t)$ the aggregate arrival rate at the
bottleneck after the model's delay alignment. In the discrete-time simulator, the queue then evolves as
\begin{equation}
\label{eq:model-queue}
q(t+\Delta)
=
\left[
q(t)+\big((1-p(t))y(t)-C\big)\Delta
\right]_0^B .
\end{equation}

Here, $\Delta=10~\mu$s is the simulation time step, $B$ is the buffer limit, and $[x]_0^B \triangleq \min\{B,\max\{0,x\}\}$ denotes clipping. The Classic dropping probability is $p(t)=(p'(t))^2$, following~\eqref{eq:classic-prob}. The factor $(1-p(t))$ approximates the fraction of traffic admitted into the queue, following~\cite{bbr2-fluid-model-imc-22}.

The PI2 controller updates $p'(t)$ based on the queueing delay $d(t)=q(t)/C$ and its change. While Linux DualPI2 uses the maximum of the Classic and L4S queueing delays, our Classic-only model uses only the Classic delay. This approximates coexistence when Classic delay dominates.

We use the default Linux PI2 update interval $T_{\mathrm{upd}}=16$~ms. Let $d_k=d(t_k)$ denote the queueing delay at update $k$ and $d^\star$ the PI2 delay target. The controller updates $p'$ as
\begin{equation}
\label{eq:pi2-update}
p'_{k+1}
=
\left[
p'_k
+
\alpha_{\mathrm{PI2}}(d_k-d^\star)
+
\beta_{\mathrm{PI2}}(d_k-d_{k-1})
\right]_0^1,
\end{equation}
where $\alpha_{\mathrm{PI2}}$ and $\beta_{\mathrm{PI2}}$ are the PI2 gains for the queueing-delay error and its change, respectively, and $[\cdot]_0^1$ denotes clipping to $[0,1]$. The model uses the same default PI2 gains as Linux, and $p'$ remains constant between updates. Since the PI2 control law is identical across congestion controls, differences in $p'(t)$ arise from their interactions with the queue and PI2 feedback. Each scenario is simulated for 40~s, with results averaged over 20--40~s to exclude the initial transient.

\medskip
\noindent\textit{Modeling scope.}
The model excludes Prague and the L4S queue; measured $p'$ differs little with and without Prague in the high-$p'$ regime. We use it to explain $p'$ regimes and relate them to Prague throughput through DualPI2 coupling in Section~\ref{sec:prague-model}.

\begin{figure*}[t]
    \centering

    \begin{subfigure}[t]{0.49\textwidth}
        \centering
        \includegraphics[width=\linewidth]{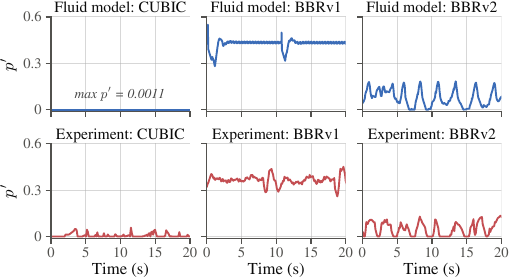}
        \caption{Representative $p'(t)$ traces in the fluid model and Linux
        ($C=100$~Mbps, base RTT $=50$~ms, PI2 target $=15$~ms).}
        \label{fig:model-traces}
    \end{subfigure}
    \hfill
    \begin{subfigure}[t]{0.49\textwidth}
        \centering
        \includegraphics[width=\linewidth]{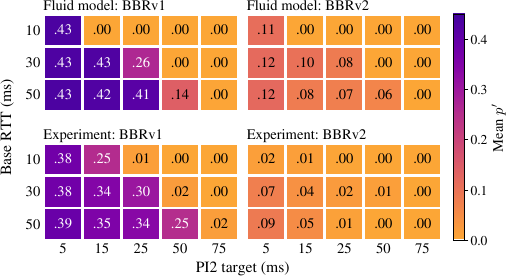}
        \caption{Mean $p'$ across the RTT/PI2-target sweep at 100~Mbps.}
        \label{fig:model-sweep}
    \end{subfigure}

    \caption{Comparison of the Classic-side PI2 signal in Linux and the
    PI2-extended fluid model. (a) The model reproduces the qualitatively
    distinct low-, intermediate-, and high-$p'$ behavior of CUBIC, BBRv2,
    and BBRv1. (b) The model captures the broad high-to-low-\(p'\) transitions, although their magnitude and location differ from Linux in some configurations.}
    \label{fig:model-comparison}
\end{figure*}

\subsection{Operating Regimes and Comparison with Linux}
\label{sec:model-comparison}

Figure~\ref{fig:model-comparison} compares the fluid model's PI2 operating regimes with Linux (Figures~\ref{fig:model-traces} and~\ref{fig:model-sweep}), using five Classic flows without Prague to isolate the PI2 behavior.

\textit{BBRv1: Persistent high $p'$.}
BBRv1's periodic bandwidth probes raise its maximum-filtered
bandwidth estimate, sustaining an offered load above the
bottleneck capacity despite losses. In the model,
BBRv1 sends about 22\% more traffic than it delivers,
forcing PI2 to keep a high dropping probability to
control the queue. At a 5~ms PI2 target, the model predicts
$p'\approx0.43$ across RTTs, consistent with
the high-$p'$ regime in Linux, apart from brief
ProbeRTT dips.

\textit{BBRv2: Lower, oscillating $p'$.}
Unlike BBRv1, BBRv2 responds to packet losses by reducing
its inflight bounds, limiting the excess load. In the model,
these reductions occur only above a 2\% loss rate
($p_C=(p')^2$, so $p'>\sqrt{0.02}\approx0.14$). As PI2 raises $p'$, losses eventually
trigger these reductions, allowing the queue and $p'$ to
decrease. Subsequent bandwidth probes increase the load
again, producing the oscillations in
Figure~\ref{fig:model-traces}. Linux BBRv2 responds to
smaller losses and uses independently timed probes,
resulting in generally lower $p'$ and less regular
oscillations. Lowering the model's loss threshold from 2\% to 0.5\% reduces mean $p'$ by a factor of $3.0$--$3.5$ (100~Mbps, 30~ms RTT, 5 and 15~ms targets), confirming that this threshold contributes to the model's higher $p'$.

\textit{Dependence on the PI2 target.}
BBRv1's CWND target is about twice its estimated
BDP, allowing one BDP in the path and another in the queue.
It can therefore sustain a queue of roughly one base RTT. PI2 must keep $p'$ high while its target is
below this delay, and $p'$ approaches zero once the target
exceeds it, so $p'$ drops at larger targets as RTT
increases. To verify that the CWND limit determines where $p'$ drops, we increase BBRv1's CWND gain from 2 to 3 in the model. This lets BBRv1 sustain an additional BDP of queued data, so $p'$ drops only once the PI2 target exceeds about two base RTTs, rather than one. BBRv2 has a similar CWND limit, so its $p'$ also approaches zero beyond about one base RTT, but its loss-responsive inflight bounds keep $p'$ lower at smaller targets.

\textit{Differences between Linux and the model.}
The model captures the main regimes, but sender details affect $p'$ and its
transitions. For BBRv1, it predicts higher $p'$ at tight targets ($0.43$
vs.\ $0.38$ in Linux), partly because Linux's ACK-based measurements smooth probing increases. Linux BBRv1
sustains larger queues (19, 42, and 63~ms vs.\ 11, 31, and 49~ms), consistent
with CWND allowances for packetization and ACK aggregation that the model
omits, so its high-$p'$ regime extends to larger targets. For BBRv2, the model overestimates $p'$ at smaller targets because it reacts only to losses above 2\% and misses Linux's CWND reductions after startup losses.

The CUBIC trace reveals a separate limitation: the model triggers CUBIC's loss response whenever the drop probability exceeds $10^{-6}$, even when actual losses would be rare, so modeled $p'$ stays near $10^{-3}$, whereas Linux CUBIC responds only to actual losses and maintains a higher $p'$. This limits quantitative comparisons with the modeled CUBIC baseline.

Overall, the model explains how BBR's sender behavior
shapes the PI2 operating regime. BBRv1's persistent excess traffic produces high $p'$, while BBRv2's loss-responsive inflight bounds lead to
lower, oscillating $p'$. Sensitivity experiments show that BBR's inflight limits
set the PI2 target where $p'$ drops,  while its loss response affects the magnitude of $p'$.

\medskip

\begin{figure}[t]
    \centering
    \begin{subfigure}[t]{0.32\columnwidth}
        \centering
        \includegraphics[width=\linewidth]{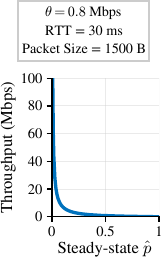}
        \caption{Prague throughput vs.\ marking $\hat{p}$.}
        \label{fig:prague-response}
    \end{subfigure}
    \hfill
    \begin{subfigure}[t]{0.66\columnwidth}
        \centering
        \includegraphics[width=\linewidth]{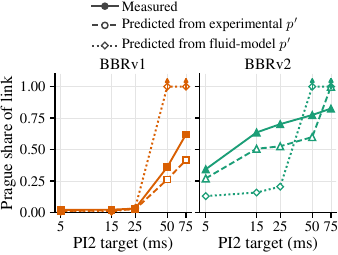}
        \caption{Measured vs.\ predicted Prague throughput (100~Mbps, 30~ms RTT).}
        \label{fig:prague-prediction}
    \end{subfigure}
    \caption{Connecting the Classic-side PI2 signal to Prague throughput.
    (a) Stronger marking sharply lowers Prague's rate.
    (b) Coupling with measured or modeled $p'$ predicts
    the degradation of Prague with BBR at tight PI2 targets.}
    \label{fig:prague-model}
\end{figure}

\section{From PI2 Signal to Prague Throughput}
\label{sec:prague-model}

To estimate Prague throughput, we use the existing
steady-state model~\cite{prague-formulation}.
Let $\hat p$ denote the marking ratio experienced by Prague,
$\tau$ its RTT, and $\tau_{\mathrm{tgt}}=25$~ms the RTT
reference used by Prague.
Then the per-flow rate in packets/second is
\begin{equation}
\label{eq:prague-throughput}
R_P(\hat p)
=
\theta_0\frac{1-\hat p}{\hat p},
\qquad
\theta_0=
\frac{2}{\max(\tau,\tau_{\mathrm{tgt}})}.
\end{equation}

We approximate the mean marking ratio from mean \(p'\) through PI2 coupling,
\begin{equation}
\label{eq:phat-coupling}
    \hat p = \min(2p',1),
\end{equation}
excluding native L4S marking. We scale the predicted
rate by five flows and 1500-byte packets, then
normalize by capacity.

Figure~\ref{fig:prague-response} shows that Prague
throughput decreases rapidly as the marking ratio increases.
At RTTs above 25~ms, the model corresponds to a CWND of approximately $2(1-\hat p)/\hat p$ packets.
For BBRv1's high $p'$ ($0.30$--$0.38$), the coupled
marking probability reaches $60$--$76\%$, reducing
Prague's predicted window to roughly one packet and
explaining its near-starvation.

For a given marking ratio and RTT, Prague's predicted
absolute throughput is independent of bottleneck capacity.
This explains the capacity dependence observed in
Figure~\ref{fig:prague-throughput}. When capacity increases
from 100 to 500~Mbps, CUBIC's measured $p'$ decreases
by approximately a factor of five at tight PI2 targets,
allowing Prague to maintain its normalized share.
With BBR, $p'$ changes much less, so Prague throughput
does not scale proportionally with capacity, reducing
its share. Similarly, with high $p'$, larger RTTs reduce Prague's predicted throughput, particularly above its 25~ms reference.

Figure~\ref{fig:prague-prediction} compares measured Prague throughput at
100~Mbps and 30~ms RTT with two predictions. The dashed curves use $p'$
measured in Linux, isolating the coupling and Prague model from fluid-model
errors. With BBRv1, they capture Prague's near-starvation at tight targets
and its recovery as the target increases: measured $p'$ drops from $0.30$ to
$0.07$ between the 25 and 50~ms targets, once the target exceeds the queue
BBRv1 sustains (Section~\ref{sec:model-comparison}). With BBRv2, the prediction follows the throughput trend but
underestimates the measured rate. Across the tested configurations,
the predictions generally identify which settings yield higher
or lower Prague throughput, but often underestimate its magnitude.
This is consistent with~\cite{prague-formulation}, which reports
a larger fitted throughput coefficient ($\theta_0$) than the theoretical
value used here.

The dotted curves instead use $p'$ from the Classic-only fluid model,
evaluating the full analytical chain. They reproduce Prague's near-starvation
with BBRv1 at tight targets, but underestimate BBRv2 throughput more severely,
since the model overestimates $p'$ (Section~\ref{sec:model-comparison}).

Some predictions at larger PI2 targets (50 and 75~ms) exceed
capacity (upward
arrows in Figure~\ref{fig:prague-prediction}) for two reasons. First, the
Classic-only model omits Prague's effect on the Classic queue: in Linux, BBRv1
alone yields $p'\leq0.02$ at large targets, but $p'$ stays around
$0.04$--$0.08$ with Prague (similarly for BBRv2), as the Classic queue
exceeds the PI2 target more often. Second, our $p'$-based throughput prediction omits
native L4S marking and capacity sharing. At a 75~ms
target with BBRv2, Prague becomes largely
capacity-limited rather than coupled-marking-limited.
In Linux, Prague obtains 83\% of the link, with
70\% of its CE marks from the native step marker.
Consequently, even the measured-$p'$ prediction exceeds capacity. Overall, the chain explains
suppression; larger targets require modeling Prague's
effect on $p'$, capacity sharing, and native L4S marking. 

\begin{figure}[t]
    \centering
    \begin{subfigure}[t]{\columnwidth}
        \centering
        \includegraphics[width=\linewidth]{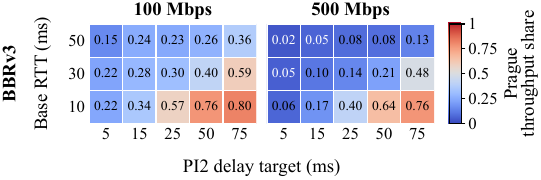}
        \caption{Normalized Prague throughput.}
        \label{fig:bbrv3-throughput}
    \end{subfigure}

    \vspace{0.5em}
    \begin{subfigure}[t]{\columnwidth}
        \centering
        \includegraphics[width=0.84\linewidth]{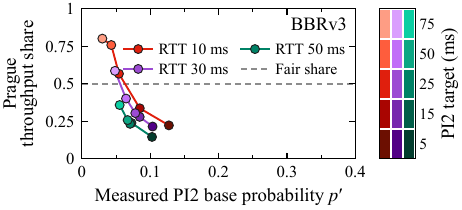}
        \caption{Normalized Prague throughput vs.\ measured $p'$ at 100~Mbps.}
        \label{fig:bbrv3-p}
    \end{subfigure}
    \caption{Five Prague and five BBRv3 flows under DualPI2, in the same setup and format as Figures~\ref{fig:prague-throughput} and~\ref{fig:p-vs-throughput}.}
    \label{fig:bbrv3}
\end{figure}

\section{Does the Problem Persist with BBRv3?}
\label{sec:bbrv3}

The fluid model of~\cite{bbr2-fluid-model-imc-22} does not include mechanisms specific to BBRv3. However, we can empirically investigate the effect of BBRv3 on the DualPI2 AQM.
We repeat the experiments described in
Section~\ref{sec:setup}, with five BBRv3 flows in the
Classic queue.

Figure~\ref{fig:bbrv3-throughput} shows that
Prague throughput degrades with BBRv3 across most
configurations, particularly at small PI2 targets.
For example, at a 30~ms RTT and 15~ms PI2 target,
Prague gets only 51~Mbps at a 500~Mbps bottleneck,
while CUBIC would get 320~Mbps.
Overall, BBRv3 degrades Prague more than BBRv2 but
generally less than BBRv1
(Figure~\ref{fig:prague-throughput}).
As the PI2 target increases, Prague's throughput
share generally improves, while the L4S queue
remains shallow (p95 delay below 1~ms).

Similar to BBRv1 and BBRv2, smaller PI2 targets
lead to higher $p'$ and lower Prague throughput
with BBRv3 (Figure~\ref{fig:bbrv3-p}).
BBRv3 generally produces higher $p'$ than BBRv2
but lower than BBRv1
(Figures~\ref{fig:p-vs-throughput} and~\ref{fig:bbrv3-p}).
At a 5~ms PI2 target and 100~Mbps, $p'$ reaches
$0.10$--$0.13$.
Like BBRv2, BBRv3 reduces its inflight bounds in
response to loss, avoiding BBRv1's persistent
high-$p'$ regime. However, BBRv2 can exit
bandwidth probing early based on an inflight
threshold, whereas BBRv3 continues probing until either
loss exceeds its threshold, or the estimated
bandwidth saturates~\cite{bbrv3-ietf117}.
This behavior is consistent with BBRv3's higher
measured $p'$ compared with BBRv2.

\section{Discussion and Future Work}
\label{sec:discussion}

Our experiments use a controlled dumbbell topology
with five long-lived flows per class and an 8-BDP buffer,
allowing us to compare Classic congestion controls
under consistent conditions. Future work will evaluate
the observed behavior with different flow counts,
more diverse application traffic, buffer sizes, and
more realistic network topologies, including heterogeneous
RTTs and multiple bottlenecks.

Our analytical framework aims to explain the
mechanisms behind the observed behavior rather
than predict exact values. The fluid model simplifies Linux BBRv1/BBRv2
behavior and omits the Prague sender, DualPI2's
L4S queue, and its scheduler. As a result, it does not capture L4S flows'
effect on $p'$, native L4S queue marking, or capacity
sharing, limiting throughput prediction accuracy
particularly in scenarios where coupled marking
is not dominant. Nevertheless, the model captures the key differences
between BBRv1's loss-insensitive behavior and BBRv2's
loss-responsive inflight control, explaining their
distinct $p'$ regimes and the observed Prague throughput
degradation when coupled marking dominates. It also
allows us to isolate the effects of individual BBR
mechanisms by varying their parameters, as demonstrated
in Section~\ref{sec:model-comparison}.

Our investigation of BBRv3 with DualPI2 is strictly experimental, because the fluid model of~\cite{bbr2-fluid-model-imc-22} only captures BBRv1 and BBRv2 behavior. 
Future work will extend the model to BBRv3 and
incorporate Prague and the L4S queue.

Finally, the scope of this paper is limited to characterizing and explaining the coexistence problem, which had previously been observed~\cite{anrw-L4S,L4S-PAM} but not fully explained. In future work, we may investigate mechanisms to mitigate this problem, to facilitate the safe deployment of L4S.

\section{Conclusion}
\label{sec:conclusion}

We characterized TCP Prague coexistence with CUBIC, BBRv1, BBRv2, and BBRv3 under
DualPI2 across bottleneck capacities, RTTs, and PI2 delay targets. BBR can
drive the PI2 base probability $p'$ far above CUBIC levels, and through
coupling, this elevated $p'$ suppresses Prague throughput even while L4S
queueing delay stays low. With our PI2 extension, the fluid model~\cite{bbr2-fluid-model-imc-22} explains BBRv1's
persistent high-$p'$ regime, BBRv2's lower, oscillating $p'$, and how the high-to-low-$p'$ transition depends on the PI2 target relative to RTT. By
combining it with DualPI2 coupling and the Prague throughput
model~\cite{prague-formulation}, we capture severe Prague suppression at tight
PI2 targets, though less accurately near and beyond the transition. Experiments with BBRv3 show that the problem is
not limited to older BBR versions.

The L4S architecture promises to maintain low latency even under heavy network load, without
sacrificing throughput. However, the first large-scale measurements of its performance in the wild~\cite{feamster} found that while many real-time application flows see a substantial improvement in latency, others see no benefit or degraded performance. This suggests that L4S performance in practical deployments may be affected by coexistence issues, such as the ones described here, and that addressing these may improve L4S deployability.

\section*{Acknowledgment}
This work was supported by the New York State Center for Advanced Technology in Telecommunications (CATT), NYU WIRELESS, and the NSF under award 2552015.

\bibliographystyle{IEEEtran}
\bibliography{references}

\end{document}